\documentclass[pra,showpacs,aps,amsmath,amssymb,twocolumn]{revtex4-1}
\usepackage{graphicx}
\usepackage{graphics}
\usepackage{amsmath,amsfonts,amssymb}
\usepackage{bm}
\usepackage{epsfig}
\usepackage[english,french]{babel}
\usepackage[utf8]{inputenc}
\usepackage{xcolor}
\usepackage{ulem}

\begin{document}
\title{The strong field approximation within a Faddeev-like formalism for laser-matter interactions}
\author{Yu. Popov$^{1,2}$, A. Galstyan$^{3}$, F. Mota-Furtado$^{4}$, P.F. O'Mahony$^{4}$ and B. Piraux$^{3}$}
\affiliation{$^{1}$Skobeltsyn Institute of Nuclear Physics, Moscow State University, Moscow, Russia\\ 
        $^{2}$Joint Institute for Nuclear Research, Dubna, Russia\\ 
        $^{3}$Institute of Condensed Matter and Nanosciences, Universit\'e catholique de Louvain, 2 chemin du cyclotron, Box L7.01.07, B-1348 Louvain-la-Neuve, Belgium\\ 
        $^{4}$Department of Mathematics, Royal Holloway, University of London, Egham, Surrey TW20 0EX, United Kingdom}
\begin{abstract}
We consider the interaction of atomic hydrogen with an intense laser field within the strong-field approximation. 
By using a Faddeev-like formalism, we introduce a new perturbative series in the binding potential of the 
atom. As a first test of this new approach, we calculate the electron energy spectrum in the very simple case of a 
photon energy higher than the ionisation potential. We show that by contrast to the standard perturbative series
in the binding potential obtained within the strong field approximation, the first terms of the new series converge 
rapidly towards the results we get by solving the corresponding time-dependent Schr\"odinger equation.
\end{abstract}
\maketitle

\section{Introduction}
\label{intro}
The strong field approximation (SFA) and its variants represent the main theoretical tool for treating the interaction of an atom or a molecule with a strong laser field. It is usually referred to as the Keldysh-Faisal-Reiss (KFR) theory since these three authors contributed significantly to its development, starting with Keldysh's 1965 seminal paper in which he introduced the first model that laid the foundation of our understanding of laser-atom interactions \cite{Keldysh}. This model describes, in the length gauge, the transition of a one-active electron atom from an initially unperturbed bound state to a dressed continuum state while assuming that the excited states do not play any role. The dressed continuum state is a Volkov state that describes a free electron in the presence of the oscilla\-ting field. It includes the electron-field dipole interaction at all orders while neglecting the Coulomb potential of the residual ion. Keldysh introduced the so-called adiabati\-city parameter $\gamma=\omega\sqrt{2I_p}/E$ where, in atomic units, $\omega$ is the photon energy, $I_p$ the ionisation potential and $E$ the electric field amplitude. In the limit $\gamma\ll 1$, ionisation occurs predominantly through a tunneling process while for $\gamma\gg 1$, ionisation proceeds by the absorption of several photons. Much later, Faisal \cite{Faisal}  and Reiss \cite{Reiss} developed an approach based on the S-matrix formalism in the velocity gauge.\\

In the strong field regime  when $\gamma\ll 1$, the usual perturbation theory in the electric field is not applicable. Simi\-larly, it is not possible to treat the binding potential  as a perturbation from the beginning since in that case, we cannot even generate a bound ground state. In the SFA, the binding potential is dominant until ionisation occurs whereas the strong oscillating field takes over after the ionisation \cite{Bauer}. After the ionisation the electron undergoes a quiver motion driven by the external oscillating field. Within the S-matrix formalism, the high-order SFA terms describe the multiple rescatte\-ring of the electron by the ionic core. Note that Faisal's S-matrix treatment can be developed also in the length gauge. In that case, one obtains the Keldysh result at the first order. \\

The SFA has provided valuable insight into the phy\-sical mechanisms underlying various processes such as the ionisation of atoms or molecules by strong infrared fields and high-order harmonic generation \cite{Lewenstein1,Lewenstein2} and even processes involving more than one electron like non-sequential double photo-ionisation of atoms \cite{Krausz}. However, SFA has a serious flaw: this theory is not gauge invariant raising the question of a preferable gauge depending on the physical process under consideration. Recently, we have reformulated the SFA and have shown that the SFA and its variants may be grouped into a set of families of approxi\-mation schemes \cite{sfa_paper}. We introduced an ansatz describing the electron wave packet as the sum of the initial state wave function times a phase factor and a function which is the perturbative solution  in the Coulomb potential of an inhomogeneous time-dependent equation. It is the phase factor that characterizes a given family. In each of these fa\-milies, the velocity and the length gauge version of the approximation scheme lead to the same results at each order in the binding potential. By contrast, irrespective of the gauge, approximation schemes belonging to different families give different results at each order thereby proving the non-existence of a preferable gauge. By looking for a gauge invariant formulation of his S-matrix approach, Faisal \cite{Faisal1,Faisal2} arrived at similar conclusions. \\

Our reformulation of the SFA has allowed us to gain more insight into the validity of the SFA schemes. We addressed two important questions; (i) the role of the Coulomb potential in the final state and (ii) the convergence of the perturbative series in the Coulomb potential. We showed that taking the Coulomb potential into account in the final state modifies significantly the low-energy part of the above-threshold ionisation spectrum of the ejected electrons. By considering a very simple but instructive case where the laser frequency is higher than the ionization potential in atomic hydrogen, we showed that the perturbative series in the Coulomb potential diverges. Although this is not a mathematical proof of the non-convergence of this perturbative series in the Coulomb potential, it casts some doubt on the pertinence of the interpretation of some strong field processes on the basis of the SFA scheme.\\

Here, we introduce, within the SFA, a new perturbative series in the Coulomb potential of the ionic core. This is done by using a Faddeev-like reformulation of the Time-Dependent Schr\"odinger Equation (TDSE). In this new approach, the excited states are taken into account allowing, for instance, a possible recombination of the electron at each rescattering. In order to test this approach, we treat the same simple case as before where the laser frequency is higher than the ionisation potential and calculate the electron energy spectrum. Our results obtained with the usual SFA approach and with this new approach, are compared to those we get by solving numerically the full TDSE. By contrast to the usual SFA perturbative series which diverges, the first three terms of the new series give results that are already very close to TDSE results. \\

This contribution is organized as follows. In Section 2 we initially briefly review our reformulation of the SFA. We then describe how we use our new formalism, inspired by Faddeev's approach, to generate the new perturbative series within the SFA. In Section 3 we present and discuss our results and finally in Section 4 we draw our conclusions and our perspectives on future directions.
Unless stated otherwise, atomic units combined with the gaussian unit system for the electromagnetic field are used throughout this contribution.\\

\section{Theory}
\subsection{Preliminary definitions}
Before proceeding with the theory, let us first define va\-rious important quantities. We consider an external pulsed field linearly polarized along the z-axis and assume that the dipole approximation is valid. The vector potential $\vec{A}(t)$ and the electric field $\vec{\mathcal{E}}(t)$ at a given time $t$ are defined as follows:
\begin{eqnarray}
\label{eq_field}
\frac{1}{c}\vec{A}(t)&=&-b'(t)\vec{e},\\ 
\vec{\mathcal{E}}(t)&=&b''(t)\vec{e},\\
b'(t)&=&\frac{1}{\omega_0} \sqrt{\frac{I}{I_0}}\sin^2(\pi\frac{t}{T})\sin(\omega_0t),
\end{eqnarray}
where $\vec{e}$ is a unit vector along the polarization axis that we assume to coincide with the z-axis. $\omega_0$ is the field frequency and $I$ is the pulse peak intensity, $I_0=3.5\times10^{16}$ W/cm$^2$ being the atomic unit of intensity. The pulse va\-nishes for $t\leq 0$ and $t\geq T$ where $T=2\pi N/\omega_0$ is the total pulse duration and $N$, the total number of optical cycles within the pulse. It is also convenient to define the following field related expression:
\begin{equation}
\label{eq_field_zeta}
\zeta(t)=\frac{1}{2}\int_0^t \mathrm{d}\xi [b'(\xi)]^2.
\end{equation}
In the next sections, we use the so-called causal time-dependent Green's function associated to the Coulomb and the dipole interaction and defined in the configuration space. The causal Coulomb Green's function $G_c^+(\vec{r},t;\vec{r}',\xi)$ is expressed as, 
\begin{multline}
\label{eq_green_coul}
G_c^+(\vec{r},t;\vec{r}',\xi)=\\
-\mathrm{i}\theta(t-\xi)\sum_\alpha
\tilde\varphi_\alpha(\vec{r})e^{-i\varepsilon_\alpha(t-\xi)}\tilde\varphi^*_\alpha(\vec{r}'),
\end{multline}
in terms of the Coulombic bound and continuum state wave functions $\tilde\varphi_\alpha(\vec{r})$. It means that the summation over $\alpha$ becomes an integral when 
$\alpha$ refers to continuum states. The causal Green's function associated to the dipole interaction is expressed as,
\begin{multline}
\label{eq_green_volkov}
G_d^+(\vec{r},t;\vec{r}',\xi)=\\
-\mathrm{i}\theta(t-\xi)\int\frac{\mathrm{d}^3p}{(2\pi)^3}\tilde\chi_g(\vec{r},\vec{p},t)\tilde\chi_g^*(\vec{r}',\vec{p},\xi), 
\end{multline}
in terms of the Volkov wave functions $\tilde\chi_g(\vec{r},\vec{p},t)$ in a given gauge $g$ ($g\equiv V$ for the velocity gauge and $g\equiv L$ for the length gauge). In the velocity gauge, the Volkov wave function is:
\begin{equation}
\tilde\chi_V(\vec{r},\vec{p},t)=e^{i\left(\vec{p}\cdot\vec{r}-p^2t/2+b(t)(\vec{e}\cdot\vec{p})-\zeta(t)\right)}.
\end{equation}
The Volkov wave function in the length gauge can be obtained from the Volkov wave function in the velocity gauge by applying the usual G\"oppert-Mayer $(V\rightarrow L)$ gauge transformation:
\begin{equation}
\tilde\chi_L(\vec{r},\vec{p},t)=e^{-ib'(t)(\vec{e}\cdot\vec{r})}\tilde\chi_V(\vec{r},\vec{p},t),
\end{equation}
where $L$ and $V$ refer to the length and the velocity gauge respectively.

\subsection{High-order SFA theory}
In a recent paper, we reformulated the general theory behind the SFA \cite{sfa_paper}. We summarise here the main results obtained in order to make the paper self contained.  Our approach which provides a simple and consistent  way of  regrouping the  different existing SFA based theoretical schemes is briefly described in this section. \\

Depending on the gauge in which we write the TDSE, we define two families of approximation schemes.
Let us first start with the TDSE in the velocity gauge. We introduce an ansatz to express the electron wave packet in the velocity gauge:
\begin{equation}
\label{eq_phi1_V}
\tilde\Phi_V(\vec{r},t)=e^{-\mathrm{i}\varepsilon_0 t}\tilde\varphi_0(\vec{r})+\tilde F_{1,V}(\vec{r},t),
\end{equation}
and in the length gauge:
\begin{equation}
\label{eq_phi1_L}
\tilde\Phi_L(\vec{r},t)=e^{-\mathrm{i}b'(t)(\vec{e}\cdot\vec{r})-\mathrm{i}\varepsilon_0 t}\tilde\varphi_0(\vec{r})+\tilde F_{1,L}(\vec{r},t),
\end{equation}
where $\tilde\varphi_0(\vec{r})$ is the atomic hydrogen ground state wave function, the energy of which is $\varepsilon_0$. The length gauge expression for the 
electron wave packet $\tilde\Phi_L(\vec{r},t)$ is obtained by applying the G\"oppert-Mayer gauge transformation to $\tilde\Phi_V(\vec{r},t)$. As a result, expressions (9) and (10) of the electron wave packet should lead to the same values of the observables. The functions $\tilde F_{1,V}(\vec{r},t)$ and $\tilde F_{1,L}(\vec{r},t)$ are solutions of an inhomogeneous equation. After substituting expressions (9) and (10) in the same TDSE we started with ({\it i.e.} in the velocity gauge), we obtain the following inhomogeneous  equation for $\tilde F_{1,V}(\vec{r},t)$ and $\tilde F_{1,L}(\vec{r},t)$:
\begin{multline}
\label{eq_inh_tdse_vg}
\left[\mathrm{i}\frac{\partial}{\partial t}+\frac{1}{2}\triangle_r- \mathrm{i}b'(t)(\vec{e}\cdot\vec{\nabla}_r)+\frac{Z}{r}-\zeta'(t)\right] \tilde F_{1,V}(\vec{r},t) =\\
= [\mathrm{i}b'(t)
(\vec{e}\cdot\vec{\nabla}_r)+\zeta'(t)]e^{-\mathrm{i}\varepsilon_0 t}\tilde\varphi_0(\vec{r}),
\end{multline}
and,
\begin{multline}
\label{eq_inh_tdse_lg1}
\left[\mathrm{i}\frac{\partial}{\partial t}+\frac{1}{2}\triangle_r- b''(t)(\vec{e}\cdot\vec{r})+\frac{Z}{r}\right] \tilde F_{1,L}(\vec{r},t) = \\
=e^{-\mathrm{i}b'(t)(\vec{e}\cdot\vec{r})}[\mathrm{i}b'(t)(\vec{e}\cdot\vec{\nabla}_r)+\zeta'(t)]e^{-\mathrm{i}\varepsilon_0 t}\tilde\varphi_0(\vec{r}).
\end{multline}
\vspace{0.3cm}

The second family of approximation schemes is obtained by starting with the length gauge TDSE. As for the first family, we introduce an ansatz to express the electron wave packet in the length gauge:
\begin{equation}
\label{eq_phi2_L}
\tilde\Phi_L(\vec{r},t)=e^{-\mathrm{i}\varepsilon_0 t}\tilde\varphi_0(\vec{r})+\tilde F_{2,L}(\vec{r},t),
\end{equation}
and in the velocity gauge:
\begin{equation}
\label{eq_phi2_V}
\tilde\Phi_V(\vec{r},t)=e^{\mathrm{i}b'(t)(\vec{e}\cdot\vec{r})-\mathrm{i}\varepsilon_0 t}\tilde\varphi_0(\vec{r})+\tilde F_{2,V}(\vec{r},t),
\end{equation}
After substitution of these two expressions for the electron wave packet in the length and velocity gauge in the length gauge TDSE, we obtain the following inhomogeneous equations for $\tilde F_{2,L}(\vec{r},t)$ and $\tilde F_{2,V}(\vec{r},t)$:
\begin{multline}
\label{eq_inh_tdse_lg2}
\left[\mathrm{i}\frac{\partial}{\partial t}+\frac{1}{2}\triangle_r-b''(t)(\vec{e}\cdot\vec{r})+\frac{Z}{r}\right] \tilde F_{2,L}(\vec{r},t) = \\
=b''(t)e^{-\mathrm{i}\varepsilon_0 t}(\vec{e}\cdot\vec{r}){\tilde\varphi}_0(\vec{r}).
\end{multline}
and
\begin{multline}
\label{eq_inh_tdse_vg2}
\left[\mathrm{i}\frac{\partial}{\partial t}+\frac{1}{2}\triangle_r- \mathrm{i}b'(t)(\vec{e}\cdot\vec{\nabla}_r)+\frac{Z}{r}-\zeta'(t)\right] \tilde F_{2,V}(\vec{r},t) = \\
=b''(t)e^{\mathrm{i}b'(t)(\vec{e}\cdot\vec{r})
-\mathrm{i}\varepsilon_0 t}(\vec{e}\cdot\vec{r})\tilde\varphi_0(\vec{r}).
\end{multline}

\noindent
The inhomogeneous equations (\ref{eq_inh_tdse_vg}) and (\ref{eq_inh_tdse_lg2}) can be solved ana\-lytically at order zero in the Coulomb potential, there\-by providing the corresponding wave packet at the first order. This wave function can then be projected on the different continuum waves or on bound states. We can also calculate the norm of the wave function, and, consequently, the ionisation yield within the SFA.\\

To conclude this brief description of the theory behind SFA, we would like to stress the following points:\\
\begin{itemize}
\item The electron wave packets belonging to the same fa\-mily lead by construction to the same result at each order for the observables. This is not the case for electron wave packets belonging to different families where the terms of the perturbative expansion differ at each order. However, the total sum of all orders gives the same results irrespective of the family. The fact that gauge invariance only holds at each order within a given family shows that it is meaningless to talk about a preferable gauge. Whether or not there is a preferable family of schemes, is still an open question.\\
\item The functions  $\tilde F_{i,g}(\vec r,t)$ (with $i=1,2$ and $g$ referring to the gauge) are the solutions of inhomogeneous differential equations. This means that they have to be normalized. However, in the limit of very low ionisation,  the norm is approximately one.\\
\item This approach allows one not only to get the analytical result for SFA electron wave packets, but it provides also a tractable way to  generate numerically a perturbative series in the Coulomb potential for the electron wave packet, by solving iteratively the inhomogeneous equations. This numerical approach helped us to study in detail the role the long range of the Coulomb potential in the output channel and to analyze the convergence of the perturbative series in one particular case.\\
\end{itemize}

The SFA has been very useful to elucidate the me\-chanism underlying various strong field phenomena. In particular, it allowed to unveil the physical processes that lead to high-order harmonic generation by atoms exposed to low frequency fields \cite{Lewenstein2}. However, beside the problem of the gauge and the possible divergence of the  SFA series expansion in the Coulomb potential which were discussed in \cite{sfa_paper}, there are also questions regarding one of the basic assumption of the SFA namely the fact that the role of the excited bound states is normally ignored in the SFA. It is important to remember that originally, the SFA was supposed to be valid in the case of negative ions which involve short range binding potentials. Perelomov, Popov and Terent'ev (PPT) \cite{Per1,Per2,Per3} developed a model based on the approximate solution in the length gauge of a time-dependent Lippmann-Schwinger like equation involving a Green's function similar to Eq. (6) in the length gauge, and expressed in terms of the classical action. For electrons bound by short range potentials, their approach reproduces the Keldysh theory. In the limit where the Coulomb field can be regarded as a small perturbation compared to the external field, effects due to the long range of the Coulomb potential are accounted for in the classical action. But, it is only recently that a first attempt to include the contribution of a few excited states within the SFA has been done \cite{Sereb}. Below in our new approach, we treat in a consistent way, the effect of all excited bound states through the causal Coulomb Green's function that enters  the Faddeev-like formalism described in the next section.

\subsection{Faddeev-like formalism}
\subsubsection{General equations}
We start with the TDSE without specifying the gauge for the time being:
\begin{equation}
\label{eq_gen_TDSE}
\left[i\frac{\partial}{\partial t}-H(t)\right]|\Psi(t)\rangle=0,
\end{equation}
where
\begin{equation}
H(t)=H_0+V_c+V_d(t).
\end{equation}
$H_0$ is the kinetic energy operator,  $V_c$ is the Coulomb potential and $V_d(t)$, the dipole interaction potential. The initial condition is $|\Psi(t=0)\rangle=|\varphi_0\rangle$ with $|\varphi_0\rangle$ being the atomic hydrogen ground state. Taking into account the fact that we have two potentials, we write, inspired by Faddeev's  approach,  
$|\Psi(t)\rangle=|\Psi_1(t)\rangle+|\Psi_2(t)\rangle$ and replace Eq. (17) by the two equations:
\begin{eqnarray}
\left[i\frac{\partial}{\partial t}-H_0\right]|\Psi_1(t)\rangle&=&V_c|\Psi(t)\rangle, \quad
|\Psi_1(0)\rangle=|\varphi_0\rangle, \\
\left[i\frac{\partial}{\partial
t}-H_0\right]|\Psi_2(t)\rangle&=&V_d(t)|\Psi(t)\rangle, \quad |\Psi_2(0)\rangle=0,
\end{eqnarray}
which can be rewritten as follows:
\begin{eqnarray}
\label{eq_system1}
& &\left[i\frac{\partial}{\partial t}-H_0-V_c\right]|\Psi_1(t)\rangle=V_c|\Psi_2(t)\rangle, \\
\label{eq_system2}
& &\left[i\frac{\partial}{\partial t}-H_0-V_d(t)\right]|\Psi_2(t)\rangle=V_d(t)|\Psi_1(t)\rangle,
\end{eqnarray}
with the same initial conditions. By using the Green's functions given by  Eqs. (5) and (6), we can look for the solutions of Eqs. (\ref{eq_system1}) and (\ref{eq_system2}) in terms of a perturbative expansion. At the first order, we have:
\begin{eqnarray}
\label{eq_fad_phi1}
|\Psi_1(t)\rangle&=& e^{-\mathrm{i}\varepsilon_0t}|\varphi_0\rangle +\int_0^\infty\mathrm{d}\xi G^+_c(t,\xi)V_c|\Psi_2(\xi)\rangle,\\
\label{eq_fad_phi2}
|\Psi_2(t)\rangle&=&\int_0^\infty\mathrm{d}\xi G^+_d(t,\xi)V_d(\xi)|\Psi_1(\xi)\rangle.
\end{eqnarray}
Contrary to what was done in the case of the SFA, the free term  in Eq. (\ref{eq_fad_phi1}) appears automatically and does not result from the introduction of any ansatz.  Note that both Green's functions alternate in the successive high-order terms. In addition, the presence of the Coulomb Green's function allows us to include different intermediate Cou\-lomb transitions within the perturbative scheme. In order to compare with the SFA perturbative expansion in $V_c$, we write $|\Psi(t)\rangle$ at the first order as:
\begin{eqnarray}
|\Psi(t)\rangle= &&e^{-\mathrm{i}\varepsilon_0t}|\varphi_0\rangle+\nonumber\\
&&\int_0^\infty\mathrm{d}\xi G^+_d(t,\xi)V_d(\xi)|e^{-\mathrm{i}\varepsilon_0t}|\varphi_0\rangle+\cdots,
\end{eqnarray}
in such a way that the first two terms of the above expansion coincide with the first order term of the SFA perturbative expansion.

\subsubsection{Gauge invariance}

In Eq. (18), we have two options for the expression of the potential $V_d(t)$ depending on the gauge:
\begin{equation}
V_d^{(V)}(t) = ib'(t)(\vec e\vec\cdot\vec\nabla_r)+\zeta'(t),
\end{equation}
in the velocity gauge and
\begin{equation}
V_d^{(L)}(t)=b''(t)(\vec e\cdot\vec r),
\end{equation}
in the length gauge. In order to have a gauge invariant theory for the state vectors $|\Psi_{1,2}^{(V)}(t)\rangle$ and $|\Psi_{1,2}^{(L)}(t)\rangle$, they should be related through the G\"oppert-Mayer transformation. However, the presence of the free term in Eq. (\ref{eq_fad_phi1}) which is the same in both gauges, prevents this relation to exist.\\

In order to solve this gauge problem, we show in the following that like in our theoretical treatment of the SFA, we can define two families  of approximation schemes. Here, we follow ideas of Faisal \cite{Faisal1,Faisal2}. Let us first rewrite Eq. (\ref{eq_gen_TDSE}) in two ways. Starting with the $V$-gauge, we write:
\begin{equation}
\left[i\frac{\partial}{\partial t}-H_0-V_c-V^{(V)}_d(t)\right]|\Psi^{(V)}(t)\rangle=0,
\end{equation}
and for the length gauge,
\begin{multline}
\;\;\;\;\;\;\;\;\;\;\;\left[i\frac{\partial}{\partial t}-(H_0+\Delta H^{(L)}(t))-V_c-\right.\\
\left.-(V^{(L)}_d(t)-\Delta H^{(L)}(t))\right]|\Psi^{(L)}(t)\rangle=0.
\end{multline}
Eq. (29) is valid for any $\Delta H^{(L)}(t)$. However, if we consider gauge-invariant families, we choose:
\begin{equation}
\Delta H^{(L)}(t)= b''(t)(\vec e\cdot\vec r)-ib'(t)(\vec{e}\vec\cdot\vec\nabla_r)+\zeta'(t).
\end{equation}
Let us write
\begin{multline}
\;\;H_0^{(L)}(t)=H_0+\Delta H^{(L)}(t)=\\
=-\frac12\triangle_r +
b''(t)(\vec e\cdot\vec r)-ib'(t)(\vec
e\vec\cdot\vec\nabla_r)+\zeta'(t).
\end{multline}
We introduce the velocity and the length form of the dipole interaction potential  in the first family:
\begin{equation}
V_d^{(V1)}(t)\equiv V_d^{(V)}(t) =\mathrm{i}b'(t)(\vec e\vec\cdot\vec\nabla_r)+\zeta'(t),
\end{equation}
\begin{eqnarray}
V_d^{(L1)}(t)& =& V^{(L)}_d(t)-\Delta H^{(L)}(t)\nonumber\\
&=&\mathrm{i}b'(t)(\vec e\vec\cdot\vec\nabla_r)-\zeta'(t).
\end{eqnarray}
We can now write the pair of equations satisfied by the state vectors $|\Psi_{1,2}(t)\rangle$ of the first family. For the velocity gauge, we have: 
\begin{equation}
\left[i\frac{\partial}{\partial t}-H_0-V_c\right]|\Psi_1^{(V1)}(t)\rangle=V_c|\Psi_2^{(V1)}(t)\rangle,
\end{equation}
\begin{multline}
\;\;\;\;\left[i\frac{\partial}{\partial t}-H_0-V_d^{(V1)}(t)\right]|\Psi_2^{(V1)}(t)\rangle=\\
=V_d^{(V1)}(t)|\Psi_1^{(V1)}(t)\rangle,
\end{multline}
with the initial conditions:
\begin{equation}
\label{eq_fad_fam11}
|\Psi_1^{(V1)}(0)\rangle=|\varphi_0\rangle; \quad |\Psi_2^{(V1)}(0)\rangle=0.\nonumber
\end{equation}
For the length gauge, we have:
\begin{equation}
\left[i\frac{\partial}{\partial t}-H_0^{(L)}(t)-V_c\right]|\Psi_1^{(L1)}(t)\rangle=V_c|\Psi_2^{(L1)}(t)\rangle,
\end{equation}
\begin{multline}
\left[i\frac{\partial}{\partial t}-H_0^{(L)}(t)-V_d^{(L1)}(t)\right]|\Psi_2^{(L1)}(t)\rangle=\\
=V_d^{(L1)}(t)|\Psi_1^{(L1)}(t)\rangle,
\end{multline}
with the initial conditions:
\begin{equation}
\label{eq_fad_fam12}
|\Psi_1^{(L1)}(0)\rangle=|\varphi_0\rangle;\quad |\Psi_2^{(L1)}(0)\rangle=0.\nonumber
\end{equation}
The state vectors $|\Psi_{1,2}^{(L1,V1)}(t)\rangle$ are gauge partners. It means that $|\Psi_{1,2}^{(L1)}(t)\rangle$ is obtained by applying the G\"oppert-Mayer transformation on $|\Psi_{1,2}^{(V1)}(t)\rangle$. It is important to note that the spectrum of the Hamiltonian in the lhs of Eq. (36) is coulombic. This means that all spectral Coulomb wave  functions (bound and continuum states) must be multiplied by the $L\rightarrow V$ G\"oppert-Mayer transformation factor.\\

To obtain the second family we proceed in the same way. We write:
\begin{equation}
V_d^{(L2)}(t)\equiv V_d^{(L)}(t)=b''(t)(\vec e\cdot\vec r).
\end{equation}
The two equations satisfied by the state vectors $|\Psi_{1,2}^{(L2)}\rangle$ in the length gauge, are:
\begin{equation}
\left[i\frac{\partial}{\partial t}-H_0-V_c\right]|\Psi_1^{(L2)}(t)\rangle=V_c|\Psi_2^{(L2)}(t)\rangle,
\end{equation}
\begin{multline}
\;\;\;\;\left[i\frac{\partial}{\partial t}-H_0-V_d^{(L2)}(t)\right]|\Psi_2^{(L2)}(t)\rangle=\\
=V_d^{(L2)}(t)|\Psi_1^{(L2)}(t)\rangle,
\end{multline}
with the initial conditions:
\begin{equation}
\label{eq_fad_fam21}
|\Psi_1^{(L2)}(0)\rangle=|\varphi_0\rangle;\quad |\Psi_2^{(L2)}(0)\rangle=0.\nonumber
\end{equation}
If we define
\begin{multline}
H_0^{(V)}(t)=H_0+\Delta H^{(V)}(t)=\\
=-\frac12\triangle_r -b''(t)(\vec e\cdot\vec r)+\mathrm{i}b'(t)(\vec e\vec\cdot\vec\nabla_r)+\zeta'(t),
\end{multline}
the two equations satisfied by the state vectors $|\Psi_{1,2}^{(V2)}\rangle$ in the velocity gauge, are:
\begin{equation}
\left[i\frac{\partial}{\partial t}-H_0^{(V)}(t)-V_c\right]|\Psi_1^{(V2)}(t)\rangle=V_c|\Psi_2^{(V2)}(t)\rangle,
\end{equation}
\begin{multline}
\left[i\frac{\partial}{\partial t}-H_0^{(V)}(t)-V_d^{(V2)}(t)\right]|\Psi_2^{(V2)}(t)\rangle=\\
=V_d^{(V2)}(t)|\Psi_1^{(V2)}(t)\rangle,
\end{multline}
with the initial conditions:
\begin{equation}
\label{eq_fad_fam22}
|\Psi_1^{(V2)}(0)\rangle=|\varphi_0\rangle; \quad |\Psi_2^{(V2)}(0)\rangle=0,
\end{equation}
where,
\begin{eqnarray}
V_d^{(V2)}(t)&\equiv &V_d^{(V)}(t)-\Delta H^{(V)}(t)\nonumber\\
&=&V_d^{(L)}(t)=b''(t)(\vec e\cdot\vec r).
\end{eqnarray}
As in the previous case, The spectrum of the Hamiltonian in the lhs of Eq. (42) is Coulombic so that all spectral Coulomb wave functions must be multiplied by the $V\rightarrow L$ G\"oppert-Mayer gauge transformation factor.\\

It is important to stress that within the same family, $|\Psi(t)\rangle$ evaluated in the length or the velocity gauge leads to identical results at each order for any observables. By contrast, if we compare observables calculated from $|\Psi(t)\rangle$ evaluated in each family, we obtain different results at each order but the summation of the  all order contributions leads to identical results.

\section{Results and Discussions}

\begin{figure}
\includegraphics[width=0.5\textwidth]{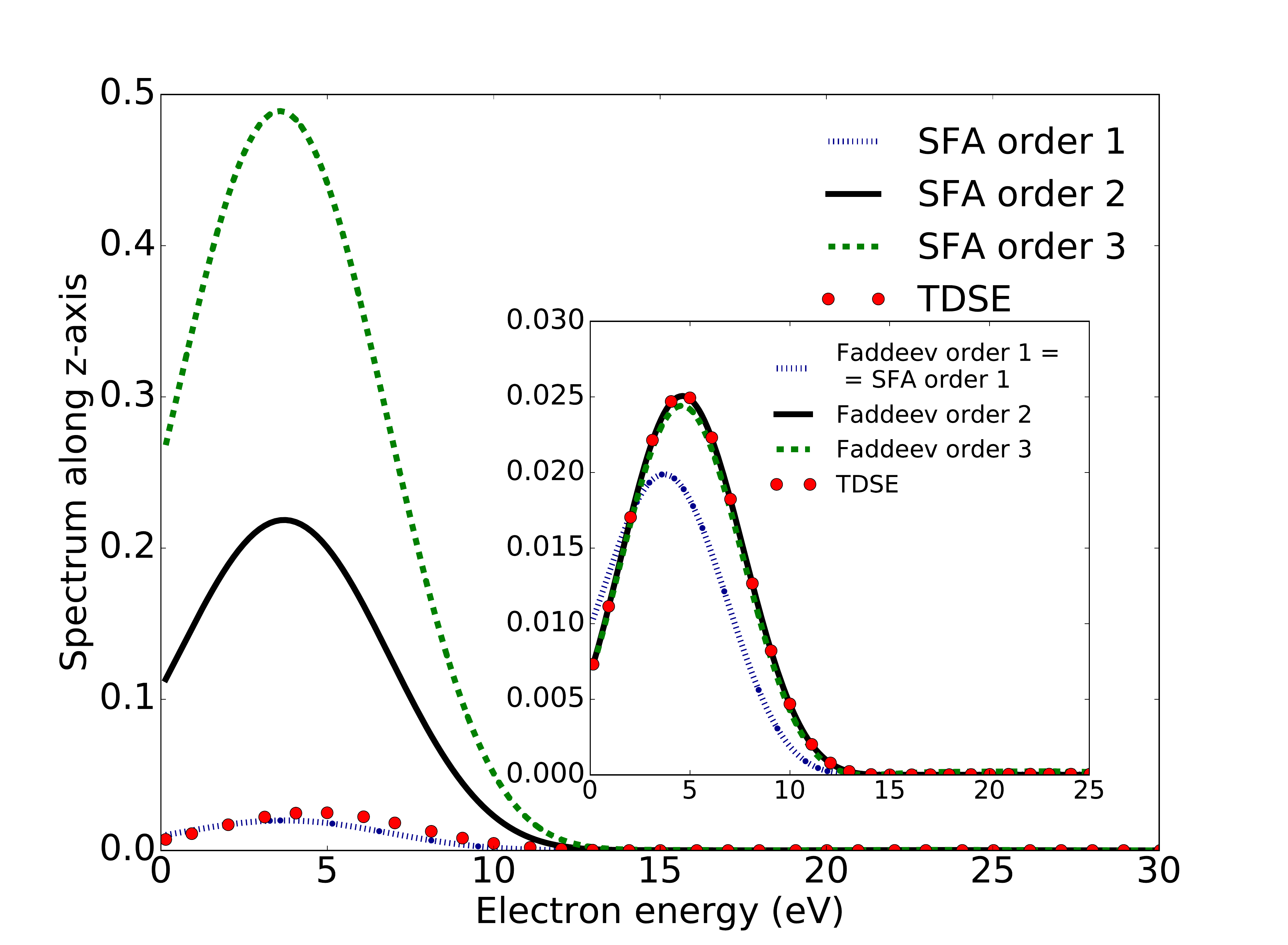}
\caption{Electron spectrum along the polarization axis resulting from the interaction of atomic hydrogen with a linearly polarized 4-cycle sine square laser pulse  of frequency $\omega=0.7$ a.u. and peak intensity $I=10^{14}$ W/cm$^2$. The results we get with the usual high-order SFA approach are compared to those, shown in the insert and obtained with our high-order SFA based Faddeev-like treatment. The vertical scale in the insert is 10 times smaller than the vertical scale of the main figure.}
\label{fig}
\end{figure}
We  consider the interaction of atomic hydrogen with a 4-cycle linearly polarized sine squared pulse of frequency $\omega=0.7$ a.u. and peak intensity $I=10^{14}$ W/cm$^2$. We calculate the electron energy spectrum along the pola\-rization axis. We have calculated the electron wave packet after the end of the interaction until the third order both within the SFA and with our Faddeev-like treatment and compared our results with the solution of the TDSE.\\

The SFA results for the electron energy spectrum have already been presented in \cite{sfa_paper} for the same field parameters. They are shown in Fig. 1 for comparison with the results based on our Faddeev-like treatment. For our SFA calculations, we used the first family of SFA schemes and the velocity gauge. We solved iteratively Eq. (11) by using a second order implicit Runge-Kutta method until up to the 15$^{th}$ order in the Coulomb potential. The results for the electron energy spectrum  are shown until the third order in Fig. 1.  They are obtained by projecting the first three orders of the SFA electron wave packet on Coulomb functions. We see in Fig. 1 that the first order SFA result is of the same order of magnitude as the TDSE result but shifted towards lower energies and with an amplitude 20\% lower (see the full blue curve on the smaller scale in the insert). We also observe that the inclusion of the second order contribution in the Coulomb potential leads to a result that is about 10 times higher than the first order result. The spectrum keeps increasing if we include the third order contribution. It is important to note that the full electron wave packet is systematically normalized to one at the end of the time propagation. Furthermore, by calculating high-order contributions to the SFA electron wave packet, we showed in \cite{sfa_paper} that the SFA perturbative series in the Coulomb potential actually diverges which manifests by the fact that the  ionisation probability tends rapidly to one after the inclusion of about 13 orders in the perturbative series.\\

We now compare the SFA results with those obtained with our Faddeev-like treatment. In these calculations, we used the first family and the velocity gauge so that at the first order, our result coincides with the first order of the SFA treatment. In our Faddeev-like treatment, the se\-cond and third order contribution to the electron wave packet $|\Psi(t)\rangle$ is obtained by solving iteratively Eqs (34) and (35) by using the same second order implicit Runge-Kutta method as in the case of the SFA. It is striking to see in the insert of Fig.1 that, within our Faddeev-like treatment, the inclusion of the second order contribution is sufficient to reproduce accurately the full TDSE results. Furthermore, the third order contribution is very small while modifying, only very slightly the second order spectrum around the maximum at 5 eV. This excellent agreement  indicates that  taking into account the excited bound states is crucial if we want to build, within the SFA, a series expansion in power of the Coulomb potential. For the present field parameters where the ionisation process is dominated by a one-photon transition, the usual SFA results, {\it i.e.} the first order term of the standard SFA expansion describes relatively  well this one-photon transition. However, the SFA high-order terms in the Coulomb potential lead to wrong results. Such terms describe electron scattering by the nucleus while neglecting the possibility of recapture into a real or virtual excited bound state. It is precisely such a process which is well described by our Faddeev-like treatment due to the presence of the causal Coulomb Green's function in the high-order terms. 

\section{Conclusions and perspectives}
In this contribution, we have studied within the SFA, the interaction of atomic hydrogen with an intense laser pulse. Our objective was to build, for the electron wave packet, a series expansion in power of the Coulomb binding potential accounting  for all excited states. We showed that this can be done in a consistent way within a Faddeev like formalism. In order to test this new approach, we considered the very simple case of a short laser pulse of frequency $\omega=0.7$ a.u. and evaluated the electron energy spectrum along the polarization axis from the electron wave packet calculated until the third order. Our results have been compared to those obtained by using the standard SFA power expansion and by solving the TDSE. In contrast to the results obtained by using the standard SFA power expansion for the electron wave packet which is known to diverge, our results obtained with our Faddeev-like formalism converge rapidly towards the TDSE results.\\

We wish to extend this approach, in particular,  to lower frequencies in the adiabatic regime. The present approach seems ideal to assess the pertinence of the recombination mechanism (frustrated tunneling) which has been proposed to explain the significant populations of excited states observed in the interaction of helium with a strong infrared pulse \cite{Nubbemeyer}. Numerically however, this calculation is much more demanding because more higher-order terms are expected to be needed to reach  convergence.

\section{Acknowledgments}
A.G. is "aspirant au Fonds de la Recherche Scientifique (F.R.S-FNRS)".
Y.P. thanks the Universit\'e Catholique de Louvain (UCL) for financially supporting several stays at the Institute of Condensed Matter and Nanosciences of the UCL. F.M.F and P.F.O'M gratefully acknowledge the European network COST (Cooperation in Science and Technology) through the Action CM1204 "XUV/X-ray light and fast ions for ultrafast chemistry" (XLIC) for financing several short term scientific missions at UCL. The present research benefited from computational resources made available on the Tier-1 supercomputer of the F\'ed\'eration Wallonie-Bruxelles funded by the R\'egion Wallonne under the grant n$^o$1117545 as well as on the supercomputer Lomonosov from Moscow State University and on the supercomputing facilities of the UCL and the Consortium des Equipements de Calcul Intensif (CECI) en F\'ed\'eration Wallonie-Bruxelles funded by the F.R.S.-FNRS under the convention 2.5020.11. A.G. and Y.P. are grateful to Russian Foundation for Basic Research (RFBR) for the financial support under the grant N14-01-00420-a. B.P. also thanks l' Agence Nationale de la Recherche fran\c{c}aise (ANR) in the context of «Investissements d'avenir» Programme IdEx Bordeaux - LAPHIA (ANR-10-IDEX-03-02).


\begin{thebibliography}{}

\bibitem{Keldysh}L.V. Keldysh, \emph{Sov. Phys. JETP} \textbf{20} (1965) 1307.
\bibitem{Faisal}F.H.M. Faisal, \emph{J. Phys. B: At. Mol. Opt. Phys} \textbf{6(4)} (1973) L89.
\bibitem{Reiss}H.R. Reiss, \emph{Phys. Rev. A} \textbf{22} (1980) 1786.
\bibitem{Bauer}G.G. Paulus and D. Bauer, in \textit{Time in Quantum Mechanics - Vol. 2} by J.G. Muga, A. Ruschhaupt, A. del Campo (Eds.) (Springer-Verlag Berlin 
                            Heidelberg 2009) 303-337.
\bibitem{Lewenstein1}M. Lewenstein, K.C. Kulander,K.J. Schafer and Ph. Bucksbaum, \emph{Phys. Rev. A}\textbf{51} (1995) 1495.
\bibitem{Lewenstein2}M. Lewenstein, Ph. Balcou, M. Yu. Ivanov, A. L'Huillier and P.B. Corkum, \emph{Phys. Rev. A}\textbf{49} (1994) 2117.
\bibitem{Krausz}F. Krausz and M. Ivanov, \emph{Rev. Mod. Phys.} \textbf{81} (2009) 163.

\bibitem{sfa_paper}Galstyan A., Chuluunbaatar O., Hamido A., Popov Yu V., Mota-Furtado F., O'Mahony P. F., Janssens N., Catoire F. and Piraux B. {\em Phys. Rev. A}     
                                    \textbf{93}, (2016) 023422.
\bibitem{Faisal1}F.H.M. Faisal, \emph{J. Phys. B: At. Mol. Opt. Phys} \textbf{40} (2007) F145.
\bibitem{Faisal2}F.H.M. Faisal, \emph{Phys. Rev. A} \textbf{75}, (2007) 063412.
\bibitem{Per1} A.M. Perelomov, V.S. Popov and M.V. Terent'ev, \emph{Sov. Phys. JETP} \textbf{23} (1966) 924.
\bibitem{Per2} A.M. Perelomov  V.S. Popov and M.V. Terent'ev, \emph{Sov. Phys. JETP} \textbf{24} (1967) 207.
\bibitem{Per3} A.M. Perelomov and V.S. Popov, \emph{Sov. Phys. JETP} \textbf{25} (1967) 336.
\bibitem{Sereb}E.E. Serebryannikov and A.M. Zheltikov, \emph{Phys. Rev. Lett.} \textbf{116} (2016) 123901.
\bibitem{Nubbemeyer} T. Nubbemeyer, K. Gorling, A. Saenz, U. Eichmann and W. Sandner, \emph{Phys. Rev. Lett.} \textbf{101} (2008) 233001.
\end{thebibliography}
\end{document}